\documentclass[a4paper,english,british,a4paper,fleqn,usenatbib]{mnras}
\usepackage[]{fontenc}
\usepackage[latin9]{inputenc}
\usepackage{url}
\usepackage{amssymb}
\usepackage{graphicx}
\usepackage[authoryear]{natbib}

\makeatletter

\providecommand{\tabularnewline}{\\}

\usepackage[T1]{fontenc}
\usepackage{ae,aecompl}

\newcommand{\kev}{keV}
\newcommand{\chandra}{\textit{Chandra}}

\newcommand{\nustar}{\textit{NuSTAR}}
\newcommand{\xmm}{\textit{XMM-Newton}}

\newcommand{\suzaku}{\textit{Suzaku}}
\newcommand{\fe}{Fe~K$\alpha$}

\newcommand{\mrk}{Mrk~335}

\title[Revealing the corona of Mrk 335]{Revealing the accretion disc corona in Mrk 335 with
  multi-epoch X-ray spectroscopy}
\author[L. Keek \& D.R. Ballantyne]{L. Keek\thanks{E-mail: l.keek@gatech.edu} and D. R. Ballantyne\\
Center for Relativistic Astrophysics, School of Physics, Georgia Institute of Technology, 837 State Street, Atlanta, GA 30332-0430, USA
}

\date{Accepted XXX. Received YYY; in original form ZZZ}
\pubyear{2015}

\makeatother

\usepackage{babel}
\begin{document}
\label{firstpage}
\pagerange{\pageref{firstpage}--\pageref{lastpage}}
\maketitle
\begin{abstract}
Active galactic nuclei host an accretion disc with an X-ray producing
corona around a supermassive black hole. In bright sources, such as
the Seyfert 1 galaxy Mrk~335, reflection of the coronal emission
off the accretion disc has been observed. Reflection produces spectral
features such as an Fe K$\alpha$ emission line, which allow for properties
of the inner accretion disc and the corona to be constrained. We perform
a multi-epoch spectral analysis of all \emph{XMM-Newton}, \emph{Suzaku},
and \emph{NuSTAR} observations of Mrk~335, and we optimize our fitting
procedure to unveil correlations between the Eddington ratio and the
spectral parameters. We find that the disc's ionization parameter
correlates strongly with the Eddington ratio: the inner disc is more
strongly ionized at higher flux. The slope of the correlation is less
steep than previously predicted. Furthermore, the cut-off of the power-law
spectrum increases in energy with the Eddington ratio, whereas the
reflection fraction exhibits a decrease. We interpret this behaviour
as geometrical changes of the corona as a function of the accretion
rate. Below $\sim10\,\%$ of the Eddington limit, the compact and
optically thick corona is located close to the inner disc, whereas
at higher accretion rates the corona is likely optically thin and
extends vertically further away from the disc surface. Furthermore,
we find a soft excess that consists of two components. In addition
to a contribution from reflection in low ionization states, a second
component is present that traces the overall flux.\end{abstract}

\section{Introduction}

\label{sect:intro} One of the most direct probes of the physics of
the inner accretion disc around an active galactic nucleus (AGN) is
through its X-ray emission. As indicated by its rapid variability,
the high-energy radiation is produced in the innermost regions of
the accretion flow, typically only several to tens of gravitational
radii ($r_{\mathrm{g}}=GM/c^{2}$, where $M$ is the mass of the black
hole and $c$ is the speed of light) from the event horizon of the
black hole \citep[e.g.,][]{gra92,mch06,uttley07,zog13}. This size
scale is $\la1(M/10^{7}M_{\odot})$~AU, which shows that it will
be impossible to image the X-ray emitting regions of AGNs for the
foreseeable future. X-ray spectroscopy is, therefore, the best tool
to test theories of accretion physics close to the central black hole.

Fortunately, many AGNs exhibit features in their X-ray spectra that
likely originate from the accretion disc \citep{fr10}. These X-ray
`reflection' features are produced by the accretion disc as a result
of its illumination from an external X-ray source, such as a magnetically
dominated corona residing above the surface of the disc \citep[e.g.,][]{grv79,hm91,hm93,hmg94,dim98,mf01}.
The strongest feature in the X-ray reflection spectrum is typically
the \fe\ emission line at 6.4~\kev\ that is due to fluorescence
in a dense and relatively cold medium \citep[e.g.,][]{bwp85,nan89,pou90,gf91,np94}.
The shape of the \fe\ line can be a powerful probe of the curvature
of space-time \citep{fab89,laor91}, leading to measurements of the
spin of the super-massive black hole \citep[e.g.,][]{br06,rf08,rey13}.
Indeed, observations over the last decade by \chandra, \xmm, \suzaku\ and
\nustar\ have uncovered several examples of relativistically broadened
\fe\ lines in the spectra of AGNs \citep[e.g.,][]{bvf03,rn03,mill07,nan07,fab09,emma11,nfw12,walton12,walton13,
ris13,Parker2014}.

Models of X-ray reflection spectra \citep[e.g.][]{rf05,garcia13}
show many features in addition to the \fe\ line that collectively
can constrain the ionization state of the illuminated surface \citep{rfy99},
often parametrized by $\xi=4\pi F/n_{\mathrm{H}}$, where $F$ is
the X-ray flux irradiating the reflector and $n_{\mathrm{H}}$ is
the hydrogen number density of the reflecting surface. Since the X-ray
emission in AGNs likely originates from the corona, a measurement
of $\xi$ may provide information on the density structure of the
disc as well as the illuminating conditions produced by the corona.
Coronal generation models suggest that the fraction of the accretion
power dissipated in the corona, $f$, can depend on the accretion
rate through the disc \citep{sr84,mf02,bp09}. Thus, there could be
a dependence of $\xi$ on the accretion rate, which, if measured,
can test ideas on the geometry of the corona and how energy flows
into it. Indeed, \citet{bmr11} considered $\alpha$-disc accretion
theory and predicted a specific relationship between $\xi$ and the
AGN Eddington ratio that depended on various disc-corona properties
such as $f$, the viscosity parameter $\alpha$ \citep{ss73}, and
the black hole spin, $a$. Therefore, if $\xi$ can be estimated for
a range of Eddington ratios from a number of AGNs, significant insight
into the workings of AGN coronae could be obtained.

\citet{bmr11} examined measurements of $\xi$ culled from the literature
and showed that, although there is significant scatter, in general
objects with larger Eddington ratios tend to exhibit ionized inner
accretion discs. While suggestive, there was too much potential object-to-object
variability in the data to draw strong conclusions from the result
presented by \citet{bmr11}. A significant improvement can be made
by determining $\xi$ for several observations of a single AGN, as
this should reduce the scatter and allow for a clearer interpretation
of any relationship with $\xi$. Fortunately, the \xmm, \suzaku,
and \nustar\ archives now contain many AGNs that have multiple observations
with which to perform such an experiment. This paper presents the
first results of this project by analysing 12 X-ray spectra of the
Seyfert 1 galaxy \mrk\ that span nearly a factor of $10$ in flux.
This AGN has several properties that make it an ideal first target
for this investigation: it has been observed several times by \xmm,
\suzaku\ and \nustar\ with 3 observations exceeding 100~ks of
exposure time \citep{Gondoin2002,Crummy2006,Larsson2008,grupe08,Grupe2012,Parker2014,Gallo2015},
is known to have a relativistically broadened \fe\ line \citep[e.g.,][]{wg15},
an Eddington ratio that can approach unity\citep{vf07,vf09}, and
a reverberation-mapped black-hole mass estimate  \citep{Peterson2004,gre12}.
As is seen below, Mrk~335 provides an excellent test case for the
predicted $\xi$--Eddington ratio correlation, and a window into the
physics of AGN coronae.

After describing the employed observations and the extraction of the
X-ray spectra from the data (Section~\ref{sec:Observations}), we
first analyse the spectra with a simple power-law model (Section~\ref{sub:Simple-power-law-fit}).
These fits indicate that spectral properties such as the shape of
the iron line correlate with flux. Detailed spectral models including
ionized reflection are used to quantify these correlations, including
the $\xi$--Eddington ratio correlation (Section~\ref{sub:Spectral-Analysis}).
Finally, we interpret the observed behaviour as changes in the coronal
optical depth and geometry as a function of the Eddington ratio (Section~\ref{sect:discuss}).

\section{Observations}

\label{sec:Observations}

\subsection{Observations and data products}

\label{sub:Observations}

\begin{table}
\caption{\label{tab:List-of-Observations}List of X-ray observations from $3$
observatories}

\begin{tabular}{ccrrr}
\hline 
 & Observatory & Obs. ID & Start date & $t_{\mathrm{exp}}(\mathrm{ks})^{\mathrm{a}}$\tabularnewline
\hline 
I & \emph{XMM-Newton} & 0101040101 & 2000-12-25 & $28$\tabularnewline
II &  & 0306870101 & 2006-01-03 & $77$\tabularnewline
III &  & 0510010701 & 2007-07-10 & $16$\tabularnewline
IV &  & 0600540601 & 2009-06-11 & $95$\tabularnewline
V &  & 0600540501 & 2009-06-13 & $70$\tabularnewline
\hline 
VI & \emph{Suzaku} & 701031010 & 2006-06-21 & $151$\tabularnewline
VII$^{\mathrm{b}}$ &  & 708016010 & 2013-06-11 & $144$\tabularnewline
VIII &  & 708016020 & 2013-06-14 & $155$\tabularnewline
\hline 
IX$^{\mathrm{b}}$ & \emph{NuSTAR} & 60001041002 & 2013-06-13 & $21$\tabularnewline
X$^{\mathrm{b}}$ &  & 60001041003 & 2013-06-13 & $22$\tabularnewline
XI &  & 60001041005 & 2013-06-25 & $93$\tabularnewline
XII &  & 80001020002 & 2014-09-20 & $69$\tabularnewline
\hline 
\end{tabular}

$^{\mathrm{a}}$ Net exposure time (excluding high-background intervals)

$^{\mathrm{b}}$ Two \emph{NuSTAR} observations overlap with \emph{Suzaku}
pointing VII
\end{table}
In order to explore ionized reflection over a wide range of source
fluxes, we employ all available X-ray observations of Mrk~335 performed
with the \emph{XMM-Newton} \citep{Jansen2001}, \emph{Suzaku} \citep{Mitsuda2007},
and \emph{NuSTAR} \citep{Harrison2013} observatories (Table~\ref{tab:List-of-Observations}).
All three observatories host a focusing telescope with grazing-incidence
mirrors and CCD imaging detectors that are sensitive around $6.4\,\mathrm{keV}$,
where the Fe~K$\alpha$ line appears as a prominent reflection feature.
With the exception of the most recent \emph{NuSTAR} pointing (XII;
see Table~\ref{tab:List-of-Observations}), all observations have
been analysed before and shown to exhibit the iron line. The combined
exposure time is $0.941\,\mathrm{Ms}$ : the most used in any detailed
X-ray spectral study of Mrk~335. \emph{NuSTAR} exposures IX and X
are fully covered by \emph{Suzaku} observation VII.

Creation of the data products for our spectral analysis is similar
for all observatories. First we apply the latest calibration to the
observation data. Next, an image and a light curve are created for
the full detector in the full instrumental band-pass. We select source
events from within a circle around the source position, and background
events from an off-source region with a similar area that is devoid
of other X-ray sources. By comparing the source to the detector light
curve, we select times without strong background emission. This is
only an issue for the \emph{XMM-Newton} observations, and reduces
their effective exposure time. Excluding the time intervals with background
flares, we extract the source and background spectra, and we generate
the accompanying response files. Neighbouring bins of the source spectra
with fewer than $25$ counts are grouped, such that $\chi^{2}$ statistics
are applicable. For most observations the source flux and the instrument's
data mode was such that pile-up was no issue, observation I being
the exception. Below we describe details of this procedure for the
different observatories.

\subsubsection{\emph{XMM-Newton} }

We use data from the pn camera \citep{Strueder2001}, which is part
of the European Photon Imaging Camera (EPIC). We do not include data
from the MOS cameras, because of their smaller effective area, nor
from the RGS, which is not sensitive in the energy band that we will
consider ($>3\,\mathrm{keV}$). Data products are created from the
observation data files using the Science Analysis System (SAS) v14.0.0.
At high count rates, multiple photons may be miscounted as one event
\citep[e.g.,][]{Ballet1999}. We check for so-called pile-up by comparing
the distribution of event types as a function of energy to the distribution
predicted by SAS task \texttt{epatplot}. Substantial pile-up is only
present in observation I, when the source exhibited a high flux and
the detector was in ``Full Window'' mode with a long CCD read-out
time. For that observation we exclude the piled-up part of the point-spread
function, which is the inner $15^{\prime\prime}$ around the source
position. The outer radius of the source extraction region is $37.5^{\prime\prime}$,
and we extract spectra in the $0.3-12\,\mathrm{keV}$ energy range.
The pn spectra typically have the highest quality among our sample,
with the ratio of the source and background count rates being $135$
to $453$ (depending mostly on the source flux).

\subsubsection{\emph{Suzaku}}

Data products for the different \emph{Suzaku} instruments are created
using the software package HEASOFT v6.16. For the X-Ray Imaging Spectrometer
(XIS) we extract spectra in the energy range $0.3-10\,\mathrm{keV}$
from a $261^{\prime\prime}$ region around the source location. We
combine the spectra of all operational front-illuminated CCDs (three
for observation VI and two for VII and VIII), and we also use the
back-illuminated CCD. The calibration of the front- and back-illuminated
CCDs produces fluxes that differ by at most a few per cent,\footnote{See the Suzaku Data Reduction Guide v5.0 at \url{http://heasarc.gsfc.nasa.gov/docs/suzaku/analysis/abc/}}
which is unnoticeable given the data quality. The ratio of source
and background count rates for the XIS spectra ranges from $3$ to
$33$.

The Hard X-ray detector (HXD) is a non-imaging instrument with a collimated
$4.5^{\circ}\times4.5^{\circ}$ field of view. It has two sensors
with a combined band-pass of $10-600\,\mathrm{keV}$. Above $40\,\mathrm{keV}$,
however, the the background dominates completely, and we only consider
data from the PIN silicon diodes in the $15-40\,\mathrm{keV}$ range.
The background is strongly dominated ($\sim95\%$) by the particle
background at the satellite.\footnotemark[\value{footnote}] We use
the simulated non-X-ray background provided by the instrument team
for the respective observations, and we employ the model of \citet{Boldt1987}\footnotemark[\value{footnote}]
to account for the small contribution from the cosmic X-ray background.
From cross-calibration of \emph{Suzaku}'s instruments,\footnotemark[\value{footnote}]
we include an enhancement factor for the PIN spectra with respect
to XIS, whose value depends on the pointing of the instruments: $1.18$
for observation VI and $1.16$ for VII and VIII.

\subsubsection{\emph{NuSTAR} }

We employ both Focal-Plane Modules: FPMA and FPMB. Using HEASOFT v6.16
and calibration files with time stamp 7/2/2015, spectra are extracted
in the energy range $3.0-50\,\mathrm{keV}$ from a source region with
radius $65^{\prime\prime}$. Whereas \emph{Suzaku}'s high energy coverage
is provided by a non-imaging instrument, \emph{NuSTAR}'s detectors
generate images for their entire band-pass. \emph{NuSTAR}'s spectra
above $10\,\mathrm{keV}$ have, therefore, a substantially lower background,
and the background can be much better constrained by using an off-source
part of the detector. The ratios of source to background count rates
are between $4$ and $10$.

\subsection{Light curves and hardness ratio}

\label{sub:Light-curves-and}

\begin{figure*}
\includegraphics{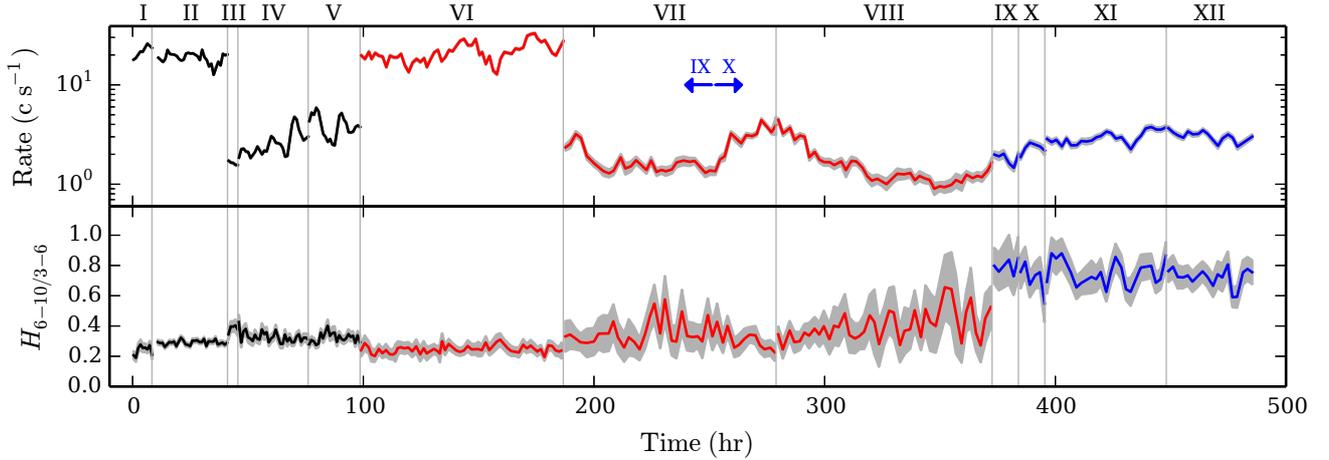}

\caption{\label{fig:lcvs}For all observations we show the count rate and hardness
ratio, $H$ ($6-10\,\mathrm{keV}$ to $3-6\,\mathrm{keV}$), at $1\,\mathrm{hr}$
(\emph{XMM-Newton} I--V) or $2\,\mathrm{hr}$ (\emph{Suzaku} VI--VIII
and \emph{NuSTAR} IX--XII) time resolution. Grey bands show the $1\sigma$
uncertainties. Two arrows indicate \emph{NuSTAR} observations that
overlap with \emph{Suzaku}. Roman numerals identify the observations
as listed in Table~\ref{tab:List-of-Observations}. They are presented
back-to-back, and are not always in time-order. Count rates are scaled
for easier comparison (see Section~\ref{sub:Light-curves-and}),
but $H$ cannot be compared between instruments because of the different
energy responses. Variations in $H$ are small in each observation
compared to the uncertainty, so the spectral shape does not evolve
significantly during one observation, and we can fit the average spectrum
in each observation.}
\end{figure*}
We investigate the time variability of the count rate and the hardness
ratio (Fig.~\ref{fig:lcvs}). For the latter we take the ratio of
the count rate in the $6-10\,\mathrm{keV}$ and $3-6\,\mathrm{keV}$
bands. We show data from \emph{XMM-Newton}'s EPIC pn, CCD 0 of \emph{Suzaku}'s
XIS, and \emph{NuSTAR}'s FPMA, which all are sensitive in the two
mentioned energy bands. Because of the different collecting areas
and energy responses of the instruments, the count rates and hardness
ratios cannot be directly compared between the different instruments.
For easier comparison of the count rates, we scale the count rates
such that they are comparable to \emph{XMM-Newton} observations II-V.
The count rate of \emph{XMM-Newton} observation I is scaled up, since
its effective area was reduced due to pile-up. \emph{Suzaku} count
rates are multiplied by a factor $14.3$ such that VI matches I, as
the two have similar X-ray fluxes (Section \ref{sub:Simple-power-law-fit}).
We scale \emph{NuSTAR} by a factor $16.7$, such that its count rates
match on average with \emph{Suzaku} during the period when observations
IX and X overlap with VII. Because of the different energy responses
of the instruments, these scaling corrections are only approximate.
For the same reason, no scaling is applied to the hardness ratios
(nor are they derived from scaled rates).

A $1\,\mathrm{hr}$ time-scale is used for \emph{XMM-Newton} and $2\,\mathrm{hr}$
for \emph{Suzaku} and \emph{NuSTAR}, because of their lower effective
areas. The light curves exhibit some time variability. The largest
variation in the count rate is by a factor $4.9$ for observation
VIII. The strongest variability is observed at lower count rates,
where the uncertainties are correspondingly larger. Most importantly,
the significance of the variations in the hardness ratio within each
observation is low. This indicates that the spectral shape does not
change substantially within each observation. This is consistent with
previous studies, that found that the average spectra of each observation
can be well fit \citep[e.g.,][]{wg15,Parker2014}. We, therefore,
refrain from splitting observations for a flux-resolved study \citep[e.g.,][]{Parker2014}.

\subsection{Bolometric Eddington ratio}

\label{sub:Bolometric-Eddington-ratio}

We are interested in how the properties of the accretion disc change
as a function of the mass accretion rate, and we use the bolometric
luminosity as a tracer of the accretion rate. From the best-fitting
model to a spectrum, the $2-10\,\mathrm{keV}$ luminosity in the rest
frame of the host galaxy is calculated using a cosmological redshift
of $z=0.025785$ \citep{Huchra1999} and assuming a flat universe
with Hubble constant $H_{0}=70\,\mathrm{km\,s^{-1}Mpc^{-1}}$ and
cosmological constant $\Lambda_{0}=0.73$. As most of the disc's thermal
emission is outside the X-ray band, we multiply the X-ray flux by
a bolometric correction factor. We use the empirical relation derived
by \citet{Zhou2010}, which provides the bolometric correction for
the unabsorbed $2-10\,\mathrm{keV}$ luminosity depending on the shape
of the continuum, specified by the photon index $\Gamma$ of the underlying
power law. Under the assumption of isotropic emission, this provides
the bolometric luminosity, $L_{\mathrm{bol}}$.

We express the luminosity as the Eddington ratio: the ratio of $L_{\mathrm{bol}}$
to the Eddington limited luminosity, $L_{\mathrm{Edd}}$. The latter
is determined as $L_{\mathrm{Edd}}=(2.1\pm0.5)\times10^{45}\,\mathrm{erg\,s^{-1}}$
using a black-hole mass of $(14\pm4)\times10^{6}\,M_{\odot}$ \citep[see also \citealt{Grier2012}]{Peterson2004}
as well as a solar hydrogen abundance of $X=0.71$.

The Eddington ratio's uncertainty is dominated by systematics. The
relation of \citet{Zhou2010} introduces a substantial systematic
uncertainty from the spread in the observations that they employed.
Including this uncertainty and the uncertainty in the Eddington luminosity,
the mean $1\sigma$ uncertainty in $\log L_{\mathrm{bol}}/L_{\mathrm{Edd}}$
is $0.77$. Without these systematic uncertainties, and only including
the uncertainties in $\Gamma$ and the $2-10\,\mathrm{keV}$ luminosity,
the mean error in $\log L_{\mathrm{bol}}/L_{\mathrm{Edd}}$ is $0.10$.
As the in-band flux is well measured, the uncertainty is mostly set
by the error in $\Gamma$. We are interested in the correlation between
parameters of a single source, so we omit the systematics from the
uncertainty reported in each data point. The systematic error would
only shift the observed trends with Eddington ratio.

\section{Simple power-law fit}

\label{sub:Simple-power-law-fit}

\begin{figure*}
\includegraphics{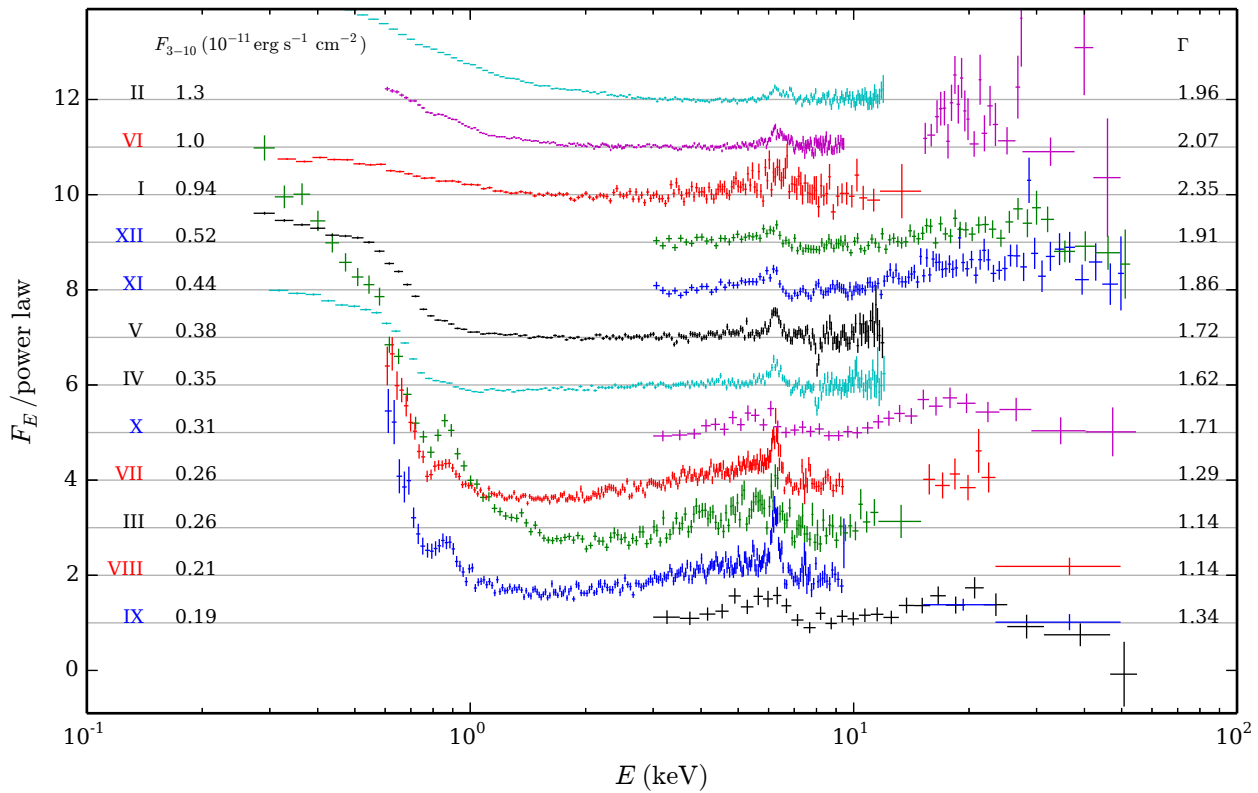}

\caption{\label{fig:po-ratios}For all observations identified with a Roman
numeral (see Table~\ref{tab:List-of-Observations}) we indicate the
$3-10\,\mathrm{keV}$ power-law flux $F_{3-10}$, the power-law index
$\Gamma$, and we show the ratio of the spectra to the best-fitting
power law model. The different observations are off set, and horizontal
lines indicate unity. We omit instruments with overlapping energies
bands: the \emph{Suzaku} XIS back-lit CCD and \emph{NuSTAR}'s FPMB.
Clearly visible are the Fe K$\alpha$ line around $E=6.4\,\mathrm{keV}$,
the Compton hump at $E\gtrsim15\,\mathrm{keV}$, as well as a soft
excess and absorption at low energies ($E\lesssim3\,\mathrm{keV}$).}
\end{figure*}
Previous analyses of these observations required a complex spectral
model that includes ionized absorption at energies $\lesssim3\,\mathrm{keV}$
as well as photoionized reflection \citep[e.g.,][]{Parker2014,wg15}.
Before we apply a similar detailed spectral model in Section~\ref{sub:Spectral-Analysis},
however, we first fit the data with a simple power law. The observed
correlation of the photon index $\Gamma$ and the X-ray flux \citep{Sarma2015},
suggests that we can gain some insight into the qualitative behaviour
of the source with a simple spectral model. Following \citet{Parker2014},
we fit a power law to the spectra in the energy intervals $3-4,\,8.5-10,$
and $40-50\,\mathrm{keV}$. This avoids absorption at low energies
as well as the Fe~K$\alpha$ line and the Compton hump, which are
prominent reflection features. These fits provide a relatively clean
look at the underlying power law. We include absorption by the Galactic
interstellar medium using the the Tübingen-Boulder model \citep{Wilms2000}
with an equivalent hydrogen column of $N_{\mathrm{H}}=3.6\times10^{20}\,\mathrm{cm^{-2}}$
\citep{Kalberla2005}, and we use solar abundances as reported by
\citet{Grevesse1998}. A cosmological redshift is applied (Section~\ref{sub:Bolometric-Eddington-ratio}).
The spectral fits are performed using XSPEC v.12.8.2 \citep{Arnaud1996},
and all presented errors in the fit parameters are at $1\sigma$.

For each spectrum we determine $\Gamma$ and the $3-10\,\mathrm{keV}$
power-law flux. We reproduce the logarithmic relation of $\Gamma$
and the X-ray flux with a break at higher fluxes \citep{Sarma2015},
which has also been observed in other AGN \citep[e.g.,][]{Perola1986,Done2000,Shih2002},
and we will quantify this relation with the detailed spectral analysis
in Section~\ref{sub:Spectral-Analysis}. The behaviour of the spectral
components that were not explicitly modelled are visualized by the
ratio of the spectra to their best-fitting power law (Fig.~\ref{fig:po-ratios}).
This provides a consistent picture of how the spectra evolve as a
function of flux. The Fe~K$\alpha$ line and the Compton hump are
visible as excesses above the power law. Considering, for example,
the \emph{Suzaku} spectra (VI--VIII), the strength of the line is
reduced at higher flux, which is expected for a more highly ionized
disc. At the low-energy side there is a strong soft excess at $E\lesssim1\,\mathrm{keV}$
as well as absorption in the $1-3\,\mathrm{keV}$ band. The relative
strength of the soft excess appears largest at low fluxes. Also, absorption
is most prominent at the lowest fluxes, whereas at the highest fluxes
the soft excess extends up to $3\,\mathrm{keV}$. Furthermore, the
data points near $50\,\mathrm{keV}$ are consistent with the power
law, except at the lowest flux values. This suggests that at low flux,
the power law has a high-energy cut-off at an energy near $\sim50\,\mathrm{keV}$,
whereas at higher flux there is no cut-off near the considered energy
band.

\section{Detailed spectral analysis}

\label{sub:Spectral-Analysis}

Although the picture sketched above is crude, it gives us the confidence
that correlations exist between various parameters and the X-ray flux,
even when ignoring the absorbed part of the spectrum ($\lesssim3\,\mathrm{keV}$;
see \citealt{Sarma2015} as well as the discussion in Section~\ref{sub:Soft-excess-and}).
Next we use a more sophisticated spectral analysis to unveil these
correlations.

First we describe the detailed spectral model, which accounts for
photoionized reflection off the disc. Next, we detail our strategy
to mitigate model degeneracies, which forces consistency in the results
of the different spectra. Finally, the fit results are presented,
and we quantify the correlations of $\log\xi$ and other model parameters
with flux.

\subsection{Spectral model}

\label{sub:Spectral-model}

The X-ray spectrum of Mrk~335 is rather complex, and consists of
several components \citep[e.g.,][]{wg15}. Thermal emission of the
accretion disc dominates the spectrum in the optical and UV. This
emission is reprocessed by Compton scattering off hot electrons in
a corona to produce a power law that extends well into the (hard)
X-ray band. In turn, the power law is reprocessed by scattering off
the photoionized accretion disc. The reflection spectrum includes
a strong fluorescent Fe K$\alpha$ emission line close to $6.4\,\mathrm{keV}$,
accompanying absorption edges at slightly higher energy, and a ``Compton
hump'' at energies $15\lesssim E\lesssim40\,\mathrm{keV}$. Furthermore,
ionized gas along the line of sight produces strong absorption below
$\lesssim3\,\mathrm{keV}$. Depending on the flux level, as many as
three separate absorption components have been detected simultaneously
in this source \citep{Longinotti2013ApJ}.

We employ the spectral model \textsc{relxill} v0.2h \citep{Garcia2014relxill,Dauser2014},
which combines \textsc{relline} \citep{Dauser2010relline} and \textsc{xillver}
\citep{garcia13}. \textsc{Xillver} provides the illuminating power
law with photon index $\Gamma$ and a high-energy exponential cut-off
at energy $E_{\mathrm{cutoff}}$. Furthermore, it includes the photoionized
reflection variant of the power law for an accretion disc with iron
abundance $A_{\mathrm{Fe}}$ relative to the solar value and ionization
parameter $\xi$. In this paper we use $\log\xi$, with $\xi$ in
units of \foreignlanguage{english}{$\mathrm{erg\,s^{-1}\,cm}$}. The
relative strength of the illumination and reflection components is
expressed as a reflection fraction, which is calculated in the $20-40\,\mathrm{keV}$
energy range. 

The second part of the \textsc{relxill} model, \textsc{relline}, describes
smoothing of the reflection spectrum due to relativistic effects,
such as gravitational redshift and relativistic Doppler broadening.
It takes into account the black-hole spin, $a$, the inclination angle
of the accretion disc with respect to the observer's line-of-sight,
and the disc's emissivity profile from an inner radius up to an outer
radius. 

A second reflection component likely originates at large distance
from the black hole, and its main contribution to the spectrum is
a narrow Fe~K$\alpha$ line. Instead of including a full reflection
model, we describe the line with a narrow unresolved Gaussian profile
at $6.4\,\mathrm{keV}$, redshifted to the reference frame of the
host galaxy. This simplified model is tested using \emph{NuSTAR} observation
XI, which has a clear detection of the narrow Fe~K$\alpha$ line
and broad energy coverage. Replacing the Gaussian line by a \textsc{xillver}
component with $\log\xi=0$, we find that outside of the Fe~K$\alpha$
line it contributes at most $13\%$ of the total flux at any given
energy. The best-fitting values of the model parameters are changed
by at most $0.3\,\sigma$, and a similar goodness of fit is obtained.

Ionized absorption is significant at energies below $\sim3\,\mathrm{keV}$,
and the simple analysis in Section~\ref{sub:Simple-power-law-fit}
indicates that correlations of the spectral properties and the X-ray
flux can be retrieved even when ignoring the absorbed part of the
spectrum (see Section~\ref{sub:Soft-excess-and} for further discussion).
Furthermore, prominent reflection features, such as the Fe~K$\alpha$
line and edge as well as the Compton hump, are at higher energies,
so the reflection spectrum can be constrained without the low-energy
part. For simplification we, therefore, ignore the spectrum below
$3\,\mathrm{keV}$, and omit spectral components for the soft excess
and ionized absorption. Absorption by an ionized gas may introduce
spectral lines above $3\,\mathrm{keV}$. Where needed, we include
narrow Gaussian profiles to fit these lines.

We apply the same cosmological redshift and absorption by the Galactic
interstellar medium as in Section~\ref{sub:Simple-power-law-fit}.
For the absorption model we use solar abundances that match those
used in calculating the \textsc{xillver} models \citep{Grevesse1998}.

\subsection{Reducing model degeneracies}

\label{sub:Reducing-degeneracies-in}

When the above-described model is applied to spectra of limited quality,
there are multiple solutions with similar goodness of fit. For example,
below $3\,\mathrm{keV}$ the uncertain nature of the soft excess and
the strong absorption pose a challenge for determining the power-law
parameters. We, therefore, restrict the fits to $E>3\,\mathrm{keV}$.
Furthermore, the ionized and the distant reflectors can be hard to
distinguish by a fit: we include a full description of the dominant
component with ionized reflection and from distant reflection only
the narrow Fe~K$\alpha$ emission line.

Another source of degeneracy in the model parameters is the Fe~K$\alpha$
line \citep[e.g.][]{Dauser2013}. Because of the strong dependence
of the line's strength, shape, and energy on the ionization state
of the disc, it is crucial for determining $\log\xi$. Its properties
are, however, also influenced by the reflection fraction, the disc's
iron abundance, $A_{\mathrm{Fe}}$, and the relativistic effects described
with \textsc{relline}. The latter depend on the the black-hole spin
$a$, the inclination angle, and the disc's emissivity profile. In
total a combination of eight parameters determines the properties
of the iron line. Consequently, multiple combinations of parameter
values produce a similar goodness of fit. Indeed, significantly different
parameter values have been obtained for individual spectra of Mrk~335
\citep[e.g.,][]{wg15}, whereas one would not expect, e.g., $A_{\mathrm{Fe}}$
or the inclination angle to change between observations.

To break these degeneracies, we constrain several parameters. We fix
the inclination angle to $30^{\circ}$, as one typically expects a
relatively small angle for Seyfert 1 galaxies (the fit results are
insensitive to the precise value). We take the disc emissivity to
decrease with the third power of the radius \citep[e.g.,][]{Miniutti2003}.
The outer radius of the disc is set to a large value of $400\,r_{\mathrm{g}}$,
and the inner radius is placed at the inner-most stable circular orbit,
which depends on $a$. We expect $a$ as well as $A_{\mathrm{Fe}}$
to be the same for all observations. Therefore, we perform a combined
fit of all \emph{XMM-Newton} spectra and tie $a$ and $A_{\mathrm{Fe}}$
between the different spectra, whereas all other free parameters are
different for each spectrum. The \emph{XMM-Newton }spectra typically
have the highest quality, and including more spectra leads to an exceedingly
problematic fitting procedure. We find $a=0.89\pm0.05$ and $A_{\mathrm{Fe}}=3.9\pm0.7$.
When fitting the \emph{Suzaku} and \emph{NuSTAR} spectra, we fix $a$
and $A_{\mathrm{Fe}}$ to these values.

This procedure substantially reduces the model degeneracies. Furthermore,
as the fixed and tied parameters have the same values across all spectra,
correlations among the other parameters are more easily retrieved.

\subsection{Fit results}

\label{sub:Fit-results}

\begin{figure}
\includegraphics{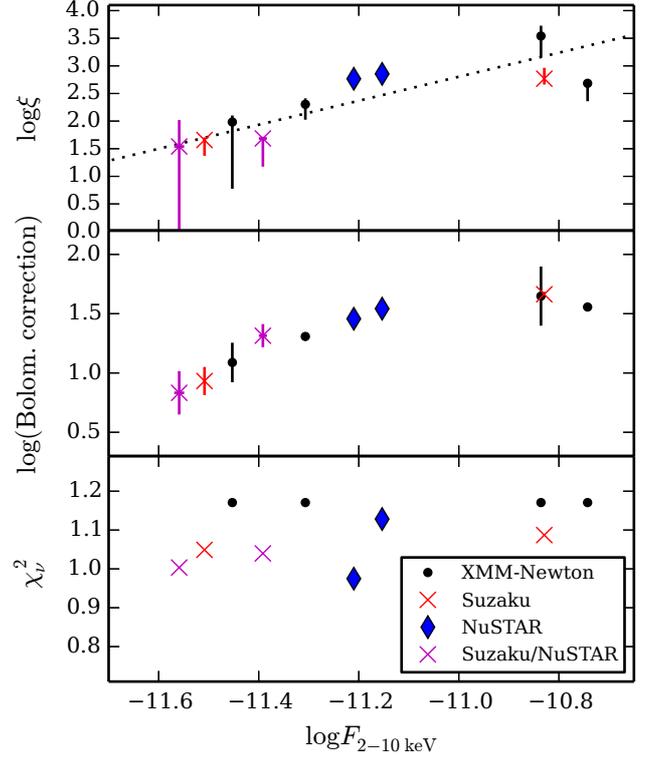}

\caption{\label{fig:f-xi}Ionization parameter $\log\xi$, bolometric correction
factor following \citet{Zhou2010}, and goodness of fit $\chi_{\nu}^{2}$
as a function of the flux in the $2-10\,\mathrm{keV}$ band, $F_{2-10\,\mathrm{keV}}$,
in units of $\mathrm{erg\,s^{-1}cm^{-2}}$. The dotted line represents
the best-fit to the data points for $\log\xi$. The observations have
the same order in flux as in Fig.~\ref{fig:po-ratios}.}
\end{figure}
Using the procedure described above, we fit all spectra. For back-to-back
observations with the same instrument where the spectral shape does
not change significantly (Fig.~\ref{fig:lcvs}), we combine the spectra.
This is the case for \emph{XMM-Newton} pointings IV and V, as well
as \emph{Suzaku} observations VII and VIII. \emph{NuSTAR}'s IX and
X were observed simultaneously with \emph{Suzaku}'s VII (Fig.~\ref{fig:lcvs}),
and we fit IX and X each together with the parts of VII that overlap
in time. The overlap is only a small part of the total exposure time
of VII. For all spectra we obtain good fits with $\chi_{\nu}^{2}\le1.17$
(Fig.~\ref{fig:f-xi}). Unresolved red-shifted Gaussian profiles
model spectral lines when needed, and we find emission lines at $(7.01\pm0.04)\,\mathrm{keV}$
(in the rest frame of the host galaxy) for II and $(7.60\pm0.05)\,\mathrm{keV}$
for VII and VIII as well as an absorption line at $(8.2501\pm0.02)\,\mathrm{keV}$
for IV and V. The presence of these lines does not significantly influence
the goodness of fit.

\begin{figure}
\includegraphics{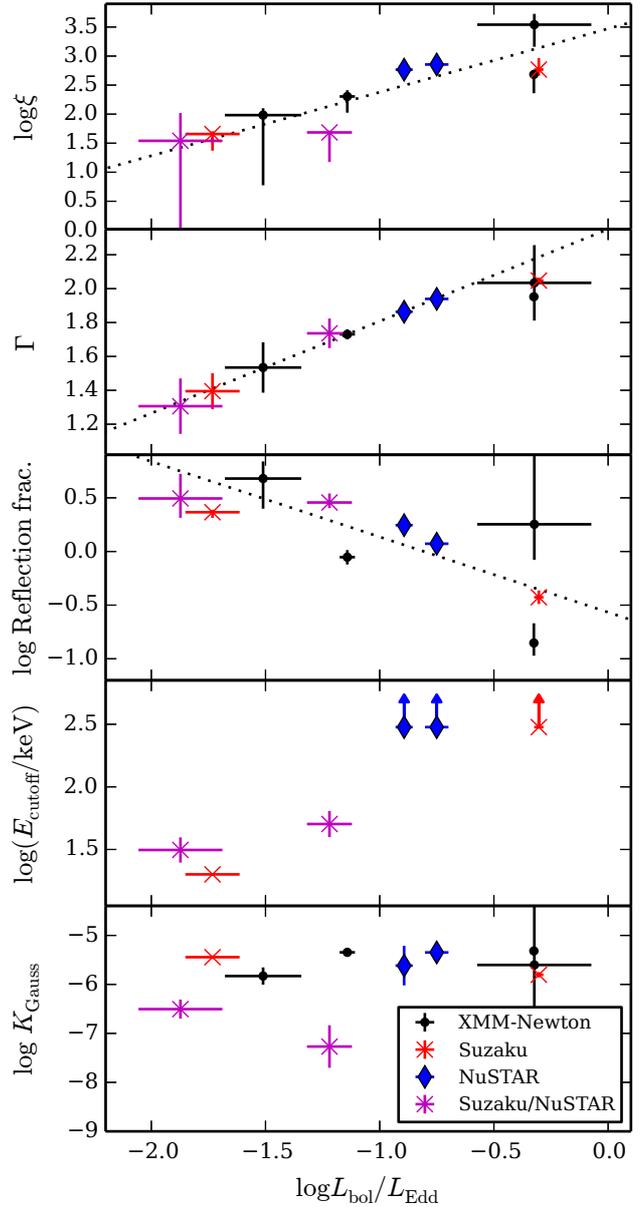}

\caption{\label{fig:ledd-xi}As a function of the bolometric Eddington ratio,
$\log L_{\mathrm{bol}}/L_{\mathrm{Edd}}$, the ionization parameter
$\log\xi$, power-law photon index $\Gamma$, the reflection fraction,
power-law cut-off energy $E_{\mathrm{cutoff}}$, and the normalization
of the narrow Fe~K$\alpha$ line from distant reflection. Dotted
lines indicate the best fits to the data points. Order of the observations
with $\log L_{\mathrm{bol}}/L_{\mathrm{Edd}}$ is the same as with
flux in Fig.~\ref{fig:f-xi}, except that $\log L_{\mathrm{bol}}/L_{\mathrm{Edd}}$
is largest for \emph{Suzaku}'s VI, followed by \emph{XMM-Newton}'s
I (with the largest errors) and II.}
\end{figure}
\begin{table*}
\caption{\label{tab:Best-fitting-values}Best fitting values of the model parameters
for fits to all spectra, $E>3\,\mathrm{keV}$ (see Section~\ref{sub:Spectral-model}
and \ref{sub:Reducing-degeneracies-in} for full details of the spectral
model). The \emph{XMM Newton} spectra are fit simultaneously, and
yield a black hole spin of $a=0.89\pm0.05$ and an iron abundance
relative to solar of $A_{\mathrm{Fe}}=3.9\pm0.7$. $A_{\mathrm{Fe}}$
and $a$ are fixed to these values in the \emph{Suzaku} and \emph{NuSTAR}
fits. \emph{NuSTAR} spectra IX and X are fit together with the overlapping
parts of \emph{Suzaku}'s VII. If $E_{\mathrm{cutoff}}$ cannot be
measured, we assume a value of $300\,\mathrm{keV}$; for VII and VIII
$E_{\mathrm{cutoff}}$ is pegged to the lower domain boundary, which
is marked with ``p.'' }

\begin{tabular}{cccccccc}
\hline 
Observation & $\Gamma$ & $\log\xi$ & $E_{\mathrm{cutoff}}$ & Reflection & $K_{\mathrm{Gauss}}$ & $\log F_{2-10\,\mathrm{keV}}$ & $\chi^{2}/\mathrm{dof}$\tabularnewline
(Table~\ref{tab:List-of-Observations}) &  & $[\log(\mathrm{erg\,s^{-1}\,cm})]$ & $[\mathrm{keV}]$ & fraction & $[10^{-6}\mathrm{c\,s^{-1}cm^{-2}}]$ & $[\log(\mathrm{erg\,s^{-1}cm^{-2}})]$ & \tabularnewline
\hline 
\emph{XMM-Newton} &  &  &  &  &  &  & \tabularnewline
I & $2.0\pm0.2$ & $3.5_{-0.4}^{+0.2}$ & $>300$ & $1.8_{-1.3}^{+5.4}$ & $2\pm12$ & $-10.836\pm0.006$ & $588/502$\tabularnewline
II & $1.95\pm0.02$ & $2.68_{-0.32}^{+0.08}$ & $>300$ & $0.14_{-0.04}^{+0.06}$ & $4.8\pm0.6$ & $-10.7427\pm0.0014$ & ``\tabularnewline
III & $1.53\pm0.14$ & $1.98_{-1.21}^{+0.11}$ & $>300$ & $5_{-3}^{+2}$ & $1.5\pm0.6$ & $-11.453\pm0.008$ & ``\tabularnewline
IV+V & $1.73\pm0.03$ & $2.30_{-0.28}^{+0.10}$ & $>300$ & $0.89_{-0.13}^{+0.13}$ & $4.5\pm0.5$ & $-11.307\pm0.002$ & ``\tabularnewline
\hline 
\emph{Suzaku} &  &  &  &  &  &  & \tabularnewline
VI & $2.05\pm0.02$ & $2.77_{-0.10}^{+0.20}$ & $>300$ & $0.37_{-0.05}^{+0.05}$ & $1.6\pm0.2$ & $-10.829\pm002$ & $1694/1558$\tabularnewline
VII+VIII & $1.39\pm0.10$ & $1.66_{-0.29}^{+0.05}$ & $20.0_{-0\mathrm{p}}^{+0.5}$ & $2.32_{-0.28}^{+0.12}$ & $3.6\pm0.3$ & $-11.508\pm0.003$ & $2271/2165$\tabularnewline
\hline 
\emph{NuSTAR} &  &  &  &  &  &  & \tabularnewline
IX (with VII) & $1.3\pm0.2$ & $1.5_{-1.5}^{+0.5}$ & $31\pm7$ & $3.1_{-1.2}^{+1.6}$ & $0.31\pm0.14$ & $-11.559\pm0.010$ & $332/331$\tabularnewline
X (with VII) & $1.74\pm0.09$ & $1.69_{-0.51}^{+0.03}$ & $50\pm12$ & $2.9_{-0.3}^{+0.6}$ & $0.05\pm0.05$ & $-11.392\pm0.008$ & $368/354$\tabularnewline
XI & $1.86\pm0.03$ & $2.77_{-0.04}^{+0.05}$ & $>300$ & $1.8_{-0.2}^{+0.2}$ & $2\pm2$ & $-11.210\pm0.004$ & $668/685$\tabularnewline
XII & $1.94\pm0.05$ & $2.85_{-0.09}^{+0.15}$ & $>300$ & $1.18_{-0.14}^{+0.15}$ & $4\pm2$ & $-11.153\pm0.004$ & $635/563$\tabularnewline
\hline 
\end{tabular}
\end{table*}
For all observations we determine the best-fitting values and $1\sigma$
uncertainties of all model parameters as well as the flux in the $2-10\,\mathrm{keV}$
band (Table~\ref{tab:Best-fitting-values}, Fig.~\ref{fig:f-xi}
and \ref{fig:ledd-xi}). We estimate the bolometric correction depending
on $\Gamma$ (Fig.~\ref{fig:f-xi}), and calculate the Eddington
ratio $L_{\mathrm{bol}}/L_{\mathrm{Edd}}$ (Fig.~\ref{fig:ledd-xi};
Section~\ref{sub:Bolometric-Eddington-ratio}). 

\begin{figure}
\includegraphics{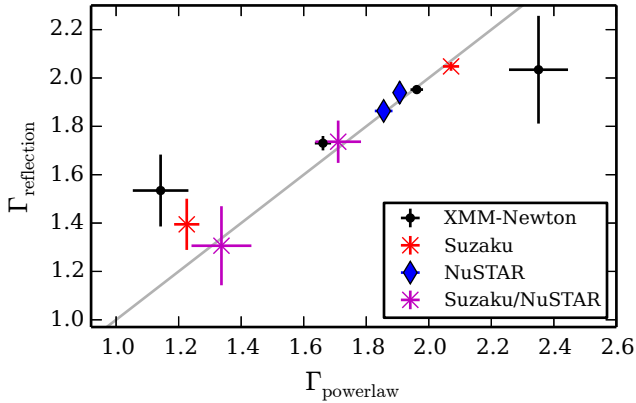}

\caption{\label{fig:gamma-gamma}Photon indices of the simple power-law fit
vs. the photoionized reflection model. Most data points are close
to the line describing $\Gamma_{\mathrm{\mathrm{power}law}}=\Gamma_{\mathrm{reflection}}$:
the power-law continuum is consistent between the simple and detailed
spectral models.}
\end{figure}
 The values of $\Gamma$ that we obtain are consistent with those
from the simple power-law fit in Section~\ref{sub:Simple-power-law-fit}
(Fig.~\ref{fig:gamma-gamma}). Importantly, this means that the simple
fits were sufficient for constraining the underlying power-law continuum,
and that the qualitative behaviour of deviations from the power law
(Fig.~\ref{fig:po-ratios}) are well described by the detailed reflection
model. Therefore, the reflection model can quantify this behaviour.
For example, the changes in the Fe~K$\alpha$ line apparent from
the power-law fits (Fig.~\ref{fig:po-ratios}) now translate into
an evolving $\log\xi$ measured with the reflection model (Fig.~\ref{fig:f-xi},
\ref{fig:ledd-xi}).

At high Eddington ratio, $\log L_{\mathrm{bol}}/L_{\mathrm{Edd}}\gtrsim-1$
(equivalent to $\log F_{2-10}\gtrsim-11.3$), no high-energy cut-off
of the power law is found, and we use a value of $E_{\mathrm{cutoff}}=300\,\mathrm{keV}$,
which is far outside the energy band of our spectra.\footnote{For spectra with a higher signal-to-noise ratio, simulations find
that $E_{\mathrm{cutoff}}$ can be measured even if it is outside
the instrument band \citep{Garcia2015}.} The fits of those spectra are insensitive to the precise value of
$E_{\mathrm{cutoff}}$: taking $E_{\mathrm{cutoff}}=1000\,\mathrm{keV}$
changes the best-fitting values of other parameters by less than $1\,\sigma$.
At lower Eddington ratios, however, we measure $E_{\mathrm{cutoff}}$
within the energy band of \emph{Suzaku} and \emph{NuSTAR}. Investigation
of confidence regions in $\chi^{2}$ space around the best fits shows
that $E_{\mathrm{cutoff}}$ is well-constrained, and does not exhibit
strong degeneracies with, e.g., the reflection fraction (see Appendix~\ref{sec:Interdependence-of-continuum}).
The lowest value, for observations VII and VIII, is at $20\,\mathrm{keV}$.
This is the lowest value provided by the \textsc{xillver} model, and
the actual value may therefore be somewhat smaller. \emph{XMM-Newton}'s
band-pass does not allow us to constrain $E_{\mathrm{cutoff}}$. This
may be an issue for the \emph{XMM-Newton} observation with the lowest
flux: III. The low flux, however, produces relatively large uncertainties,
such that any deviations in the parameter values due to an incorrect
$E_{\mathrm{cutoff}}$ are not significant.

The non-ionized reflector is likely located at a large distance from
the black hole, e.g., the outer region of the disc, a torus, or the
broad-line region \citep[e.g.,][]{Nandra2006}. The time-scale for
changes to propagate to this region is $\sim$years: similar to the
interval covered by the observations considered in our study (Table~\ref{tab:List-of-Observations}).
We, therefore, expect the normalization of the narrow Fe~K$\alpha$
line, $K_{\mathrm{Gauss}}$, to be constant. Indeed, for four observations
where $K_{\mathrm{Gauss}}$ is largest, its weighted mean is $(4.2\pm0.2)\times10^{-6}\,\mathrm{c\,s^{-1}\,cm^{-2}\,keV^{-1}}$
(Fig.~\ref{fig:ledd-xi}). For all other observations $K_{\mathrm{Gauss}}$
is lower, which may be attributed to difficulty in separating this
line from the one in the ionized reflection component. There is no
evident correlation with Eddington ratio: the correlation coefficient
is $r=0.29$ with a $p$-value for no correlation of $0.57$, and
a linear fit yields a slope that is consistent with $0$ within $1\sigma$.

\subsection{Correlations with Eddington ratio}

\label{sub:Correlations-with-Eddington}

\begin{table*}
\caption{\label{tab:Correlations-with-respect}Correlations of the ionization
parameter $\log\xi$, photon index $\Gamma$, and reflection fraction
with respect to $\log L_{\mathrm{bol}}/L_{\mathrm{Edd}}$, and for
$\log\xi$ also with respect to $\log F_{2-10}$ (first row). Provided
are the correlation coefficient $r$ and the $p$-value for the null
hypothesis (no correlation), as well as the slope, intercept, and
$\chi_{\nu}^{2}$ from a linear fit. Furthermore, we include linear
fits with an extra systematic error in the parameter measurements
such that $\chi_{\nu}^{2}\equiv1$.}

\begin{tabular}{|c|r@{\extracolsep{0pt}.}l|c|c|c|r@{\extracolsep{0pt}.}l|c|c|r@{\extracolsep{0pt}.}l|}
\hline 
Parameter & \multicolumn{2}{c|}{$r$} & $p$ & Fitted slope & Intercept & \multicolumn{2}{c|}{$\chi_{\nu}^{2}$} & $\chi_{\nu}^{2}\equiv1$ fitted slope & Intercept & \multicolumn{2}{c|}{$\sigma_{\mathrm{systematic}}$}\tabularnewline
\hline 
$\log\xi(\log F_{2-10})$ & \multicolumn{2}{c|}{$0.84$} & $2.3\times10^{-3}$ & $2.17\pm0.10$ & $26.7\pm1.1$ & \multicolumn{2}{c|}{$35$} & $1.8\pm0.4$ & $23\pm5$ & \multicolumn{2}{c|}{$13\,\%$}\tabularnewline
$\log\xi$ & \multicolumn{2}{c|}{$0.88$} & $6.9\times10^{-4}$ & $1.09\pm0.09$ & $3.47\pm0.07$ & \multicolumn{2}{c|}{$9.6$} & $1.1\pm0.2$ & $3.3\pm0.3$ & \multicolumn{2}{c|}{$11\,\%$}\tabularnewline
$\Gamma$ & \multicolumn{2}{c|}{$0.97$} & $2.1\times10^{-6}$ & $0.54\pm0.10$ & $2.35\pm0.10$ & \multicolumn{2}{c|}{$0.05$} & --- & --- & \multicolumn{2}{c|}{---}\tabularnewline
$\log$~Reflection fraction & \multicolumn{2}{c|}{$-0.73$} & $1.7\times10^{-2}$ & $-0.70\pm0.09$ & $-0.57\pm0.08$ & \multicolumn{2}{c|}{$7.7$} & $-0.1\pm0.2$ & $-0.1\pm0.2$ & \multicolumn{2}{c|}{$94\,\%$}\tabularnewline
\hline 
\end{tabular}
\end{table*}
 We quantify the correlations between the logarithm of the Eddington
ratio and $\log\xi$, $\Gamma$, and the logarithm of the reflection
fraction. For $\log\xi$ we also investigate the correlation with
the logarithm of the $2-10\,\mathrm{keV}$ flux. The correlation coefficient,
$r$, and the $p$-value for the null hypothesis (no correlation)
are determined. Furthermore, a linear function is fit to the two parameters,
taking into account the uncertainties in both, and we give the slope
and intercept in Table~\ref{tab:Correlations-with-respect}. Often
the goodness of fit, $\chi_{\nu}^{2}$, is substantially larger than
unity, due to the small errors in certain data points. In those cases
we include a systematic error, which is added in quadrature to the
uncertainties of the data points (in the ``vertical'' direction),
such that a ``perfect fit'' with $\chi_{\nu}^{2}=1$ is obtained.

$\Gamma$ correlates strongly with $\log L_{\mathrm{bol}}/L_{\mathrm{Edd}}$.
We fit a line for $\log L_{\mathrm{bol}}/L_{\mathrm{Edd}}<-0.5$ .
The data points at highest $\log L_{\mathrm{bol}}/L_{\mathrm{Edd}}$
are below this trend, which is consistent with the previously observed
break in the trend at high flux \citep[e.g.,][]{Sarma2015}.

A strong correlation is found between $\log\xi$ and the X-ray flux,
$\log F_{2-10}$, as well as with the bolometric Eddington ratio.
The $\chi_{\nu}^{2}$ values of the linear fit are, however, substantially
larger than $1$. The addition of a systematic error of $\sim12\,\%$
yields $\chi_{\nu}^{2}=1$ for both the flux and the Eddington ratio.
The obtained slope is consistent within $1\,\sigma$ with the previous
fit.  Interestingly, the slope of the linear fit is consistent within
$1\sigma$ with the measurement of \citet{bmr11}, who combined literature
values of $10$ AGNs. Also, the intercept is consistent within $2\sigma$
($1\sigma$ when including the extra systematic error; Table~\ref{tab:Correlations-with-respect}).
This suggests that the measured relation may be a universal property
of the behaviour of these sources. Our measurement is not affected
by systematic uncertainties due to differences between sources, and
yields a higher correlation coefficient. Therefore, we firmly establish
the presence of the correlation of $\log\xi$ and flux as well as
Eddington ratio for Mrk~335. 

Although the reflection fraction clearly appears to decrease for increased
values of the Eddington ratio (Fig.~\ref{fig:ledd-xi}), the correlation
is weaker because of substantial scatter in the data points. To obtain
a $\chi_{\nu}^{2}=1$ fit, a high systematic error of $94\%$ is needed,
and the correlation is no longer significant.

The cut-off energy of the power law transitions from a low value of
$\sim30\,\mathrm{keV}$ for $\log L_{\mathrm{bol}}/L_{\mathrm{Edd}}<-1$,
to $E_{\mathrm{cutoff}}\gtrsim300\,\mathrm{keV}$ at higher Eddington
ratio. The transition is, however, not sampled well enough to quantify
the correlation.

\section{Discussion}

\selectlanguage{english}%
\label{sect:discuss}\foreignlanguage{british}{ We have performed
a multi-epoch analysis of all available \emph{XMM-Newton}, \emph{Suzaku},
and \emph{NuSTAR} observations of Mrk~335 to study the behaviour
of the accretion disc corona as a function of Eddington ratio, by
inferring correlations of the flux and the spectral parameters. Detailed
models of ionized reflection can present degenerate solutions when
applied to spectra of limited quality. Other authors have found acceptable
fits to the individual spectra of Mrk~335 with alternative models,
and found values of $\Gamma$, $\xi$ \citep[e.g.,][]{wg15}, and
$E_{\mathrm{cutoff}}$ \citep[e.g.,][]{Parker2014} that differ significantly
from the trends that we uncover. We optimized our fitting procedure
explicitly to infer consistent behaviour across multiple observations
at different flux levels. This allows us to quantify strong correlations
of the Eddington ratio and $\Gamma$ \citep[see also][]{Sarma2015},
$\xi$, as well as correlations with the reflection fraction and
$E_{\mathrm{cutoff}}$. }

\selectlanguage{british}%

\subsection{The ionization parameter--Eddington ratio correlation in Mrk~335}

\label{sub:xicorr}

The previous section showed that \mrk\ exhibits a strong and statistically
significant correlation between $\xi$ and $L_{\mathrm{bol}}/L_{\mathrm{Edd}}$.
The \citet{bmr11} $\alpha$-disc prediction of this relationship,
assuming a radiation-pressure dominated accretion disc and a geometrically
thick corona (i.e., $H/R=1$, where $H$ is the height of the X-ray
source above the disc and $R$ is the radial distance along the disc\footnote{This geometry is supported by \fe\ reverberation measurements that
give the distance between the corona and the disc \citep{rm13}.}) is 
\begin{equation}
\begin{array}{c}
\xi\approx4.33\times10^{9}\left(\frac{\eta}{0.1}\right)^{-2}\left(\frac{\alpha}{0.1}\right)\left(\frac{L_{\mathrm{bol}}}{L_{\mathrm{Edd}}}\right)^{3}\left(\frac{R}{r_{\mathrm{g}}}\right)^{-7/2}\\
\\
\times R_{z}^{-2}R_{T}^{-1}R_{R}^{3}f(1-f)^{3}\,\mathrm{erg\,cm\,s^{-1}},
\end{array}\label{eq:radxi}
\end{equation}
where $\eta$ is the radiative efficiency of the disc, $f$ is the
fraction of accretion energy dissipated in the corona, $\alpha$ is
the viscosity parameter \citep{ss73}, and $(R_{R},R_{z},R_{T})$
encompass the general relativistic effects and are dimensionless functions
of $a$ and $(R/r_{\mathrm{g}})$ \citep[e.g.,][]{nt73}. The spectral
analysis of \mrk\ found good fits when the inner radius of the reflecting
region was at the ISCO of a black hole with $a=0.89$ (i.e., $R=R_{\mathrm{ISCO}}=2.39$~$r_{\mathrm{g}}$).
This removes some of the freedom present in comparing the observations
to the prediction of Eq.~\ref{eq:radxi}, and potentially eases the
physical interpretation of the results.

\begin{figure}
\includegraphics{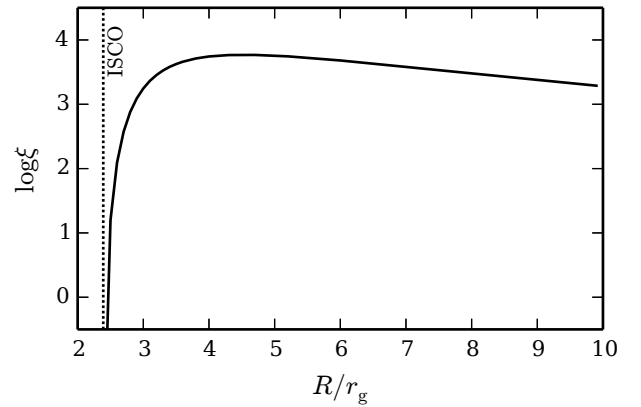}

\caption{\label{fig:xi_radius}Predicted $\log\xi$ as a function of radius,
$R$, for $f=0.45$, $a=0.89$, and $\log L_{\mathrm{bol}}/L_{\mathrm{Edd}}=-0.3$
(Eq.~\ref{eq:radxi}). $\log\xi$ drops off fast near the ISCO due
the factors $(R_{R},R_{z},R_{T})$, whereas simulations indicate that
$\log\xi$ should continue to increase due to a drop in density at
the ISCO \citep{rf08}.}

\end{figure}
However, Eq.~\ref{eq:radxi} becomes inaccurate for $R$ close to
$R_{\mathrm{ISCO}}$, as the functions $R_{z}$, $R_{T}$ and $R_{R}$
begin to diverge \citep{kro99}. This causes the disc density to increase
rapidly, and $\xi$ therefore drops precipitously at radii close to
the ISCO (Fig.~\ref{fig:xi_radius}). Such behaviour is unphysical
and is not seen in numerical simulations of accretion flows close
to the ISCO where the gas density is found to fall quickly as material
begins to plunge towards the event horizon \citep[e.g.,][]{rf08}.
Thus, unlike what is seen in Fig.~\ref{fig:xi_radius}, $\xi$ should
continue to rise as $R$ moves toward $R_{\mathrm{ISCO}}$. To correct
this issue, the run of $\xi$ with $L_{\mathrm{bol}}/L_{\mathrm{Edd}}$
is first computed at $R=4$~$r_{\mathrm{g}}$ where $(R_{R},R_{z},R_{T})$
have not begun to strongly diverge (Fig.~\ref{fig:xi_radius}), and
then this result is scaled to $R=2.4$~$r_{\mathrm{g}}$ using the
$(R/r_{\mathrm{g}})^{-7/2}$ scaling predicted from Eq.~\ref{eq:radxi}.
\begin{figure}
\includegraphics{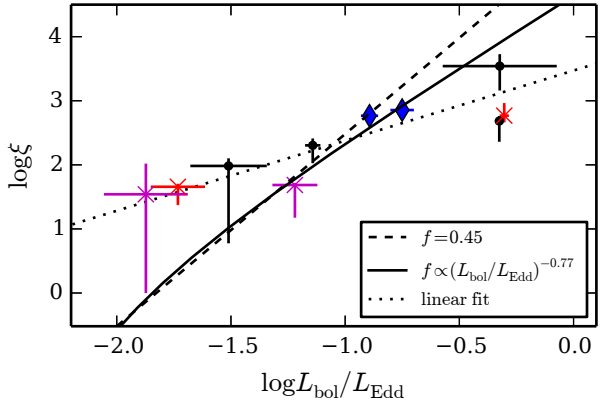}

\caption{\label{fig:xi_ledd}$\log\xi$ as a function of Eddington ratio. We
compare the data points from Fig.~\ref{fig:ledd-xi} and the best
linear fit (dotted line) to two predictions based on Eq.~\ref{eq:radxi}:
the dashed line is for $f=0.45$, and the solid line assumes $f\propto(L_{\mathrm{bol}}/L_{\mathrm{Edd}})^{-0.77}$.
See Fig.~\ref{fig:ledd-xi} for a description of the symbols.}
\end{figure}
This procedure results in the two lines seen in Fig.~\ref{fig:xi_ledd},
where the model predictions are plotted over the measurements of \mrk.
The dashed line assumes a constant $f=0.45$ \citep{vf07} and the
solid line assumes $f\propto(L_{\mathrm{bol}}/L_{\mathrm{Edd}})^{-0.77}$
\citep{sr84}. A constant $\alpha=0.1$ and $\eta=0.1$ \citep{dl11}
is also assumed for both these curves.

It is remarkable that the straightforward theoretical prediction of
Eq.~\ref{eq:radxi} provides such a decent description of the observations
of $\log\xi$ from Mrk~335. This basic level of agreement suggests
that our simple understanding of accretion discs and coronae is on
the right track. Yet, it is also clear that the theoretical slope
is too steep, with $\xi$ underpredicted at low $L_{\mathrm{bol}}/L_{\mathrm{Edd}}$
and overpredicted at large $L_{\mathrm{bol}}/L_{\mathrm{Edd}}$. This
result was also found by \citet{bmr11} using literature results,
and indicates that Eq.~\ref{eq:radxi} is missing one or more important
physical effects. \citet{bmr11} speculated how changing some of the
quantities in Eq.~\ref{eq:radxi} (e.g., $\alpha$) or the underlying
disc density may help bring the predicted $\xi$ more in agreement
with the measurements, but there were no additional constraints with
which to narrow down the possibilities. However, our spectral fits
of \mrk\
provide significant new information that may allow for a clearer picture
of the corona in this system.

\subsection{Piecing it together: a vision for the corona of Mrk~335}

\label{sub:corona} Fig.~\ref{fig:ledd-xi} shows that there is evidence
for an anti-correlation between the reflection fraction and $L_{\mathrm{bol}}/L_{\mathrm{Edd}}$
in \mrk. The strongest reflection is found when the source is in
the lowest flux state, which is consistent with light-bending enhancing
the irradiation of the disc from a compact corona \citep[e.g.,][]{mf04}.
Our spectral model indicates a disc that is strongly irradiated at
the ISCO of a spinning black hole, so, depending on the height of
the X-ray source, light bending could be important in illuminating
the disc. The enhancement of the irradiating flux due to light-bending
is not included in the predictions of Eq.~\ref{eq:radxi}, so this
is a natural explanation for why $\log\xi$ is underpredicted at low
Eddington ratios --- the flux is too low and therefore the disc is
not as ionized as observed. However, this effect would cause $\xi$
to increase at all Eddington ratios, exacerbating the overprediction
at large $L_{\mathrm{bol}}/L_{\mathrm{Edd}}$, unless there are changes
to the coronal geometry that reduce the irradiating the flux at these
Eddington ratios. The fact that the reflection fraction seems to drop
at higher values of $L_{\mathrm{bol}}/L_{\mathrm{Edd}}$ indicates
that some change in coronal geometry is occurring.

\begin{figure}
\includegraphics{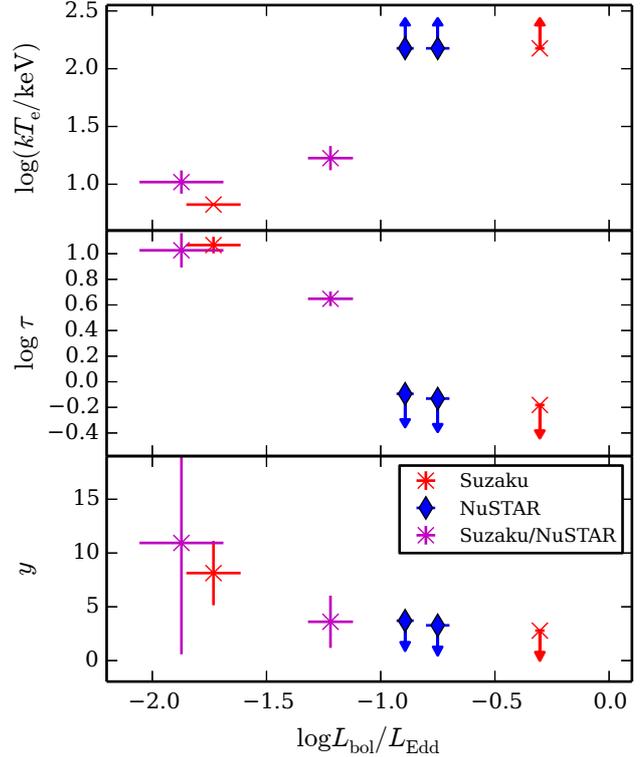}

\caption{\label{fig:ledd_compt}Properties of the Comptonizing corona: electron
temperature $kT_{\mathrm{e}}$, optical depth $\tau$, and Compton
$y$-parameter as a function of Eddington ratio.}
\end{figure}
 The trends observed in $\Gamma$ and $E_{\mathrm{cutoff}}$ (Fig.~\ref{fig:ledd-xi})
together support this interpretation. The primary X-ray power-law
is thought to be generated by thermal Comptonization in the corona
\citep[e.g.,][]{stern95,ps96,pet01}, and depends on the coronal optical
depth and temperature; e.g., 
\begin{equation}
\Gamma\simeq\left[\frac{9}{4}+\frac{m_{\mathrm{e}}c^{2}}{kT_{\mathrm{e}}\tau(1+\tau/3)}\right]^{1/2}-\frac{1}{2},\label{eq:gamma}
\end{equation}
where $\tau\gtrsim1$ is the optical depth and $kT_{\mathrm{e}}$
the electron temperature \citep[e.g.,][]{sle76}. The upper two panels
of Fig.~\ref{fig:ledd_compt} show estimates of $kT_{\mathrm{e}}$
and $\tau$ (from Eq.~\ref{eq:gamma}) as a function of $L_{\mathrm{bol}}/L_{\mathrm{Edd}}$
for \mrk. The coronal temperature is estimated using $kT_{\mathrm{e}}=E_{\mathrm{cutoff}}/3$
for $L_{\mathrm{bol}}/L_{\mathrm{Edd}}<0.1$ and $kT_{\mathrm{e}}=E_{\mathrm{cutoff}}/2$
for larger Eddington ratios \citep[e.g.,][]{pet01}. These plots clearly
suggest that the coronal optical depth diminished by about an order
of magnitude as the Eddington ratio increased in \mrk. Once the Eddington
ratio exceeded $\sim10\,\%$, the optical depth and the temperature
remained roughly constant.

One implication of this behaviour is seen in the lower panel of Fig.~\ref{fig:ledd_compt}.
Here, the Compton $y$-parameter, defined as \citep{pet01} 
\begin{equation}
y\simeq4\left(\frac{4kT_{\mathrm{e}}}{mc^{2}}\right)\left[1+\left(4kT_{\mathrm{e}}/mc^{2}\right)\right]\tau(1+\tau),\label{eq:y}
\end{equation}
is plotted versus $L_{\mathrm{bol}}/L_{\mathrm{Edd}}$. For a given
geometry, a Comptonizing corona in equilibrium will adjust so that
it has a fixed $y$ \citep[e.g.,][]{stern95}. The $y$ parameter
of the Comptonizing corona of \mrk\ changes substantially with Eddington
ratio and is only relatively constant at larger Eddington ratios (Fig.~\ref{fig:ledd_compt}).
These results strongly imply that the coronal geometry of \mrk\ evolves
from a high optical depth configuration to a lower one as the Eddington
ratio increases.

Further evidence for a change in coronal geometry can be seen by considering
the compactness of the X-ray source, which is proportional to the
ratio of the X-ray luminosity and the size of the corona \citep[e.g.,][]{gfr83,fab15}.
Considerations of energy balance and radiation processes in the corona
lead to various relationships between the compactness and $kT_{\mathrm{e}}$,
all of which point to an inverse relationship between the two; that
is, independent of the scattering processes ongoing in the corona,
the compactness is smaller for larger temperatures. Observations by
\nustar\ and other missions are now showing that multiple AGNs follow
such a trend \citep{fab15}. Assuming the radiative processes ongoing
within the corona do not change, the fact that the temperature rises
with Eddington ratio in \mrk~ implies that the compactness will
fall. Since the luminosity is also increasing with a larger $L_{\mathrm{bol}}/L_{\mathrm{Edd}}$,
then the compactness can only fall if the size scale of the radiative
source increases faster than the luminosity rise.

Putting all these pieces together paints the following portrait of
the evolving corona of \mrk. At $L_{\mathrm{bol}}/L_{\mathrm{Edd}}\la0.1$
the corona is warm, compact, optically thick and probably situated
close to the black hole. This leads to significant light bending and
illumination of the inner disc, increasing the reflection fraction
and ionization state of the reflecting region. As the Eddington ratio
increases, the corona expands and its optical depth drops, allowing
the temperature to rise. The reflection fraction also begins to fall
indicating that the disc is not being irradiated as strongly as before;
this argues that the change in coronal geometry is more in the vertical
direction away from the disc, rather than covering more disc area
which would normally increase the reflection fraction. If the X-ray
source is now further from the black hole, light bending would not
be as important, and the flux on the disc surface would be significantly
reduced. In that way, the predicted $\xi$ can be brought closer to
the observations at $L_{\mathrm{bol}}/L_{\mathrm{Edd}}\ga0.1$.

\subsection{Soft excess and absorption}

\label{sub:Soft-excess-and}

The ratio of the spectra to simple power-law fits exhibits an evolving
soft excess as well as absorption at low energies ($E<3\,\mathrm{keV}$
in Fig.~\ref{fig:po-ratios}). Previously, some studies have interpreted
the soft excess as part of the reflection signal \citep{Crummy2006},
since reflection models predict a bump in the spectrum due to bremsstrahlung
and a multitude of emission lines \citep[e.g.,][]{garcia13}. Although
the peak is at low energies, $E\lesssim10^{-2}\,\mathrm{keV}$, the
tail extends into the \emph{XMM-Newton} EPIC pn and \emph{Suzaku}
XIS bands. At large values of $\log\xi$ the feature is weak, but
at low $\log\xi$ its shape is similar to the soft excess. \citet{wg15}
used the \textsc{reflionx} reflection model \citep{rf05} to fit the
\emph{XMM-Newton} and \emph{Suzaku} spectra of Mrk~335, and included
the data at $E<3\,\mathrm{keV}$. As the X-ray count rate peaks in
the soft band, it dominates the fit, and the soft excess likely forces
small values of $\log\xi$. When they repeated their analysis with
similar reflection models as in our study, no good fits were obtained.
As these models present an improved description of reflection spectra,
it suggests that the bump at low energies is not fully described by
the reflection signal.

\begin{figure}
\includegraphics{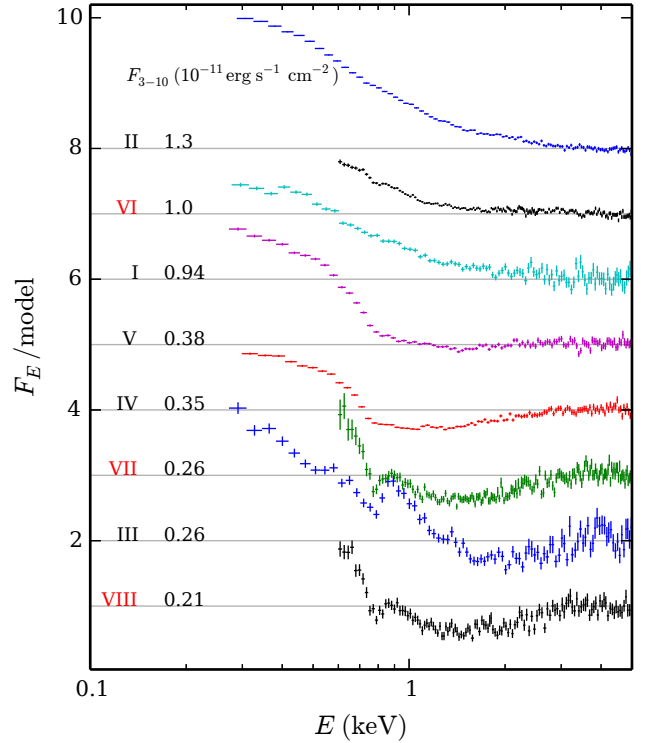}

\caption{\label{fig:refl_ratios}Ratios of \emph{XMM-Newton} and \emph{Suzaku}
spectra and the best-fitting reflection models, similar to Fig.~\ref{fig:po-ratios}
(the flux values are taken from that figure). The spectral fits were
performed for $E>3\,\mathrm{keV}$, whereas here we illustrate deviations
from the best-fitting models at lower energy due to absorption and
a soft excess.}
\end{figure}
 By ignoring the soft band in our fits, our measurement of $\log\xi$
depends more strongly on the Fe~K$\alpha$ line and (for \emph{Suzaku}
and \emph{NuSTAR}) the Compton hump. We find that $\log\xi$ correlates
with flux: as expected, stronger illumination produces a higher $\log\xi$
(Fig.~\ref{fig:f-xi}). For a cleaner view of the soft excess and
absorption, we take the ratio of the spectra to the best-fitting reflection
models from Section~\ref{sub:Spectral-Analysis} (Fig.~\ref{fig:refl_ratios}).
We show the ratio for \emph{XMM-Newton} observations down to $0.3\,\mathrm{keV}$
and for \emph{Suzaku} XIS to $0.6\,\mathrm{keV}$, whereas \emph{NuSTAR}
is not sensitive below $3\,\mathrm{keV}$. The qualitative picture
is the same as detailed in Section~\ref{sub:Simple-power-law-fit}:
absorption is stronger at low flux \citep[see also][]{Longinotti2013ApJ},
and a soft excess is present. Reflection, therefore, does not fully
account for the soft excess. Furthermore, at low flux, observations
III, VII, and VIII exhibit an additional feature just below $1\,\mathrm{keV}$,
which may signify the presence of an absorption edge. 

Previous studies found absorption columns of up to $N_{\mathrm{H}}\sim10^{23}\,\mathrm{cm^{-2}}$
\citep{Grupe2012,Longinotti2013ApJ}, which could be large enough
to impact spectral parameters such as $\Gamma$ and $\log\xi$. These
studies, however, typically did not include a reflection component,
and obtained poor fits when including $E\gtrsim2\,\mathrm{keV}$ data.
As a test, we take into account warm absorption when fitting spectrum
III, which may have relatively strong absorption as the source is
in a low-flux state. Columns of $N_{\mathrm{H}}\le10^{22}\,\mathrm{cm^{-2}}$
do not significantly change the fit results, whereas for $N_{\mathrm{H}}=10^{23}\,\mathrm{cm^{-2}}$
the predicted absorption is stronger than observed. Furthermore, \citet{Sarma2015}
analysed the \emph{XMM-Newton} spectra of Mrk~335, and obtained qualitatively
and quantitatively similar results when either fitting the $E>3\,\mathrm{keV}$
data or when including an ionized absorption component in fits of
the full instrument band. This supports our assumption. If one nevertheless
includes a strong absorption component in the spectral model, the
power law component compensates with a larger photon index. For example,
\emph{NuSTAR} observation XII exhibits a relatively high flux, where
we do not expect substantial absorption (Fig.~\ref{fig:refl_ratios}).
\citet{Wilkins2015flares} analyse this pointing in combination with
contemporaneous \emph{Swift} XRT data while including strong absorption
($N_{\mathrm{H}}\sim10^{23}\,\mathrm{cm^{-2}}$), and obtain a photon
index that is $29$ per cent higher than our best fitting value. Our
results can form the basis of an improved study into the nature of
the soft excess and ionized absorption, which may also be responsible
for some the observed spectral lines (Section~\ref{sub:Fit-results}). 

\begin{figure}
\includegraphics{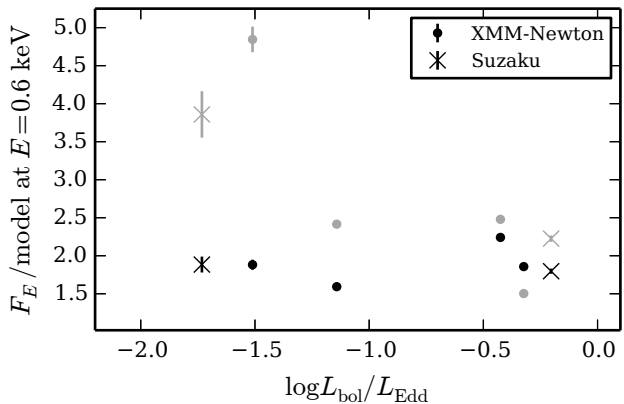}

\caption{\label{fig:refl_ratios_edd}Comparison of the ratios of the spectra
at $0.6\,\mathrm{keV}$ with respect to the reflection model (black;
Fig.~\ref{fig:refl_ratios}) and the power-law fits (grey; Fig.~\ref{fig:po-ratios})
as a function of the Eddington ratio, $\log L_{\mathrm{bol}}/L_{\mathrm{Edd}}$.
Subsequent observations (IV, V and VII, VIII) have been merged. For
clarity, the horizontal errors are omitted, and a small offset in
$\log L_{\mathrm{bol}}/L_{\mathrm{Edd}}$ is included where needed.
The large excess at low $L_{\mathrm{bol}}/L_{\mathrm{Edd}}$ for the
power law are only partially described by the reflection model: a
second component of the soft excess remains at all $L_{\mathrm{bol}}/L_{\mathrm{Edd}}$
for the reflection model.}
\end{figure}
 To quantify the strength of the soft excess we consider the ratio
of the spectra to their best-fitting models at $0.6\,\mathrm{keV}$,
which is the lowest energy that we include for \emph{Suzaku}. For
the power-law fits, the ratio is largest at low Eddington ratio, and
decreases with luminosity (grey points in Fig.~\ref{fig:refl_ratios_edd}).
This is consistent with models of reflection spectra, that predict
soft emission to be stronger for low values of $\xi$, which we find
at low Eddington ratio (Fig.~\ref{fig:ledd-xi}). Indeed, when we
take the ratio of the spectra to the best-fitting reflection models,
the ratios at low Eddington ratio are substantially reduced (black
points in Fig.~\ref{fig:refl_ratios_edd}). They are, however, not
reduced to zero, but a component remains. This component is present
at both low and high Eddington ratio, and it is relatively constant
with a mean value of $1.9$. A constant ratio implies that it is tracks
the overall flux, rather than, e.g., the reflection fraction. It may
originate from the source of seed photons for coronal X-ray emission:
the accretion disc. For example, the source could be a warm scattering
disc atmosphere \citep[e.g.,][]{Magdziarz1998,Done2012,Petrucci2013,Rozanska2015}.

\section{Conclusions}

Applying multi-epoch spectroscopy to \emph{XMM-Newton}, \emph{Suzaku},
and \emph{NuSTAR} observations of Seyfert 1 galaxy Mrk~335, we have
uncovered consistent behaviour of multiple spectral parameters as
a function of flux and, therefore, as a function of the Eddington
ratio. In particular, we find a strong correlation between the ionization
state of the inner accretion disc and the Eddington ratio. The slope
is consistent with a previous measurement for literature values of
$10$ AGNs \citep{bmr11}, suggesting that the relation describes
behaviour that is common for these sources. We firmly establish the
correlation with a higher correlation coefficient, as our result is
not affected by systematic uncertainties between sources. Furthermore,
this correlation is close to the theoretical relation predicted by
\citet{bmr11}, but the measured slope is less steep, suggesting changes
in the geometry of the corona. We infer the following behaviour of
the evolving corona. At low values of the Eddington ratio, the corona
is optically thick, compact, and close to the inner accretion disc
where light bending is strong. As the accretion rate (and the Eddington
ratio) increases, the corona expands, likely in the vertical direction.
Above $\sim10\,\%$ of the Eddington ratio, the corona is more extended,
optically thin, and light bending is less important.

Our fits to the spectra above $3\,\mathrm{keV}$ indicate that at
lower energies the soft excess consists of $2$ components. One is
due to reflection and its strength depends on the ionization state,
whereas the other does not. The strength of the latter component is
proportional to the overall flux, and possibly originates from the
accretion disc.

Compared to previous work that considered individual spectra, we find
that multi-epoch spectroscopy is essential for breaking degeneracies
in the spectral fits and for obtaining accurate spectral parameters.
Furthermore, we show that this method provides a powerful tool to
study coronal evolution. The rich archives of the previously mentioned
observatories provide the opportunity to extend this investigation
to include several other bright AGN, which will reveal whether the
behaviour that we found is common or unique to Mrk~335. Finally,
future theoretical work needs to investigate the evolution of the
coronal geometry with Eddington ratio to explain the observed behaviour
of the disc's ionization state.

\section*{Acknowledgements}

The authors acknowledge support from NASA ADAP grant NNX13AI47G and
NSF award AST 1008067.

\bibliographystyle{mnras}
\bibliography{agnrefl}

\appendix

\section{Degeneracies in the continuum parameters}

\label{sec:Interdependence-of-continuum}

\begin{figure}
\includegraphics[width=1\columnwidth]{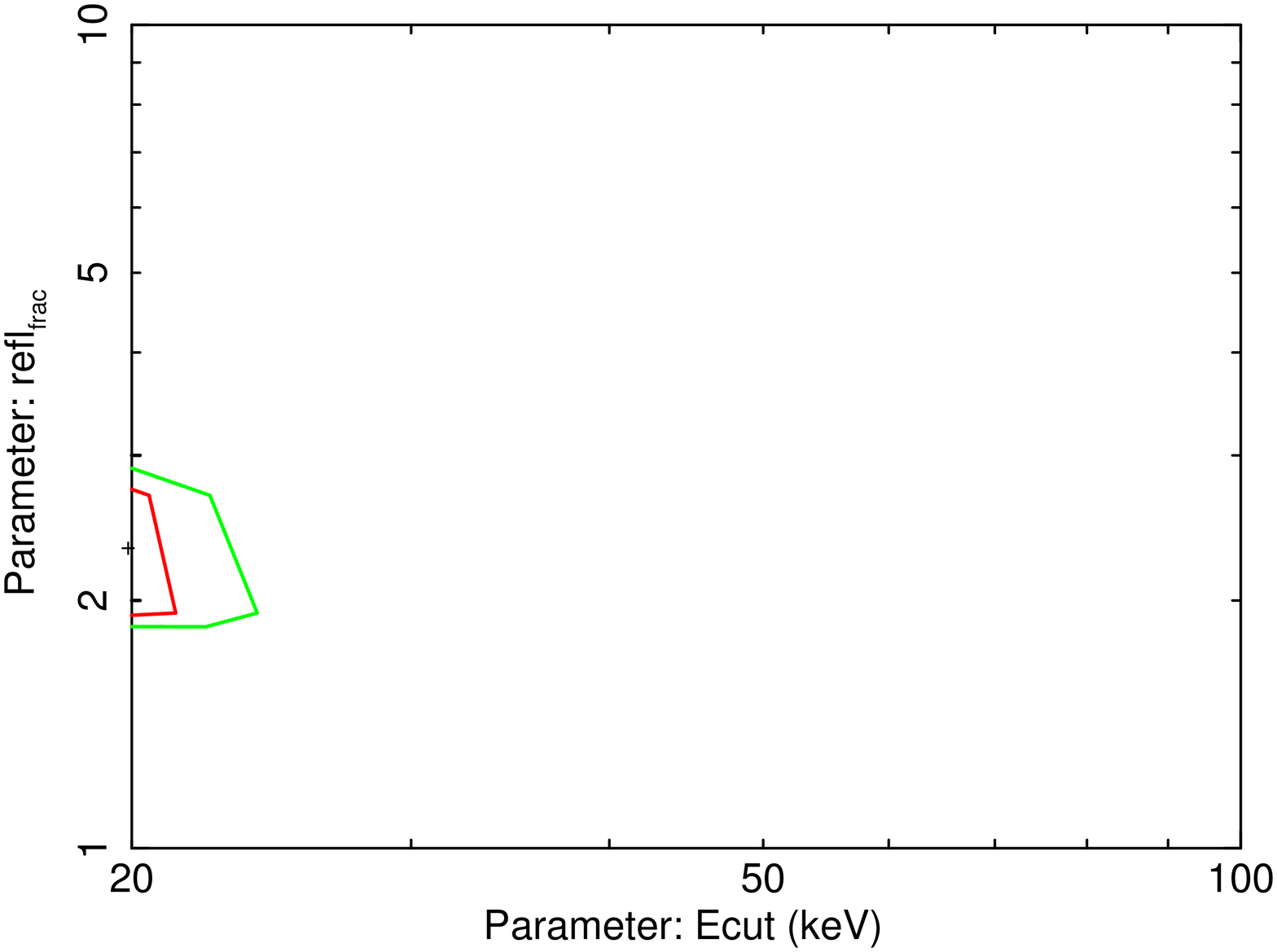} \includegraphics[width=1\columnwidth]{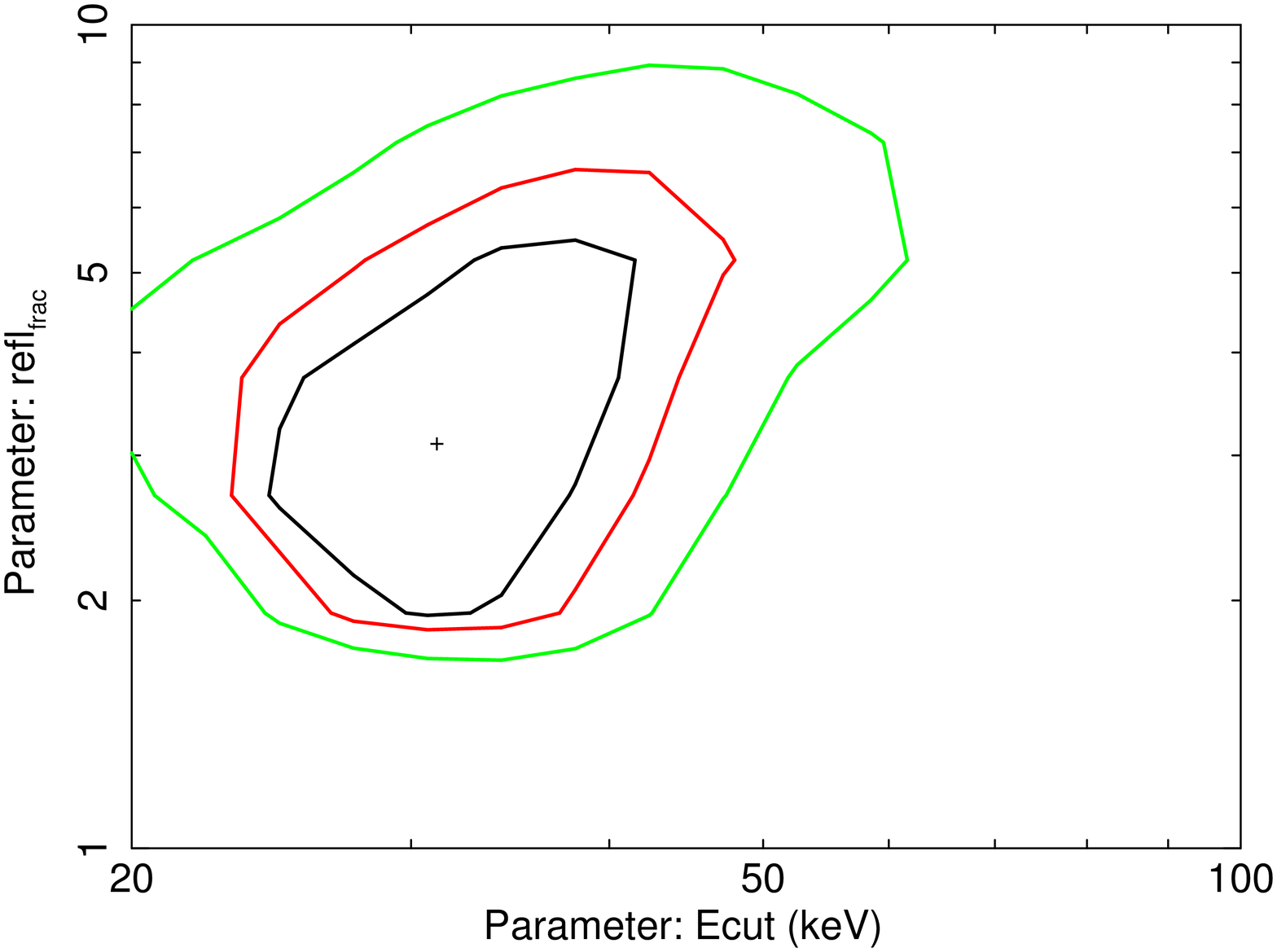}
\includegraphics[width=1\columnwidth]{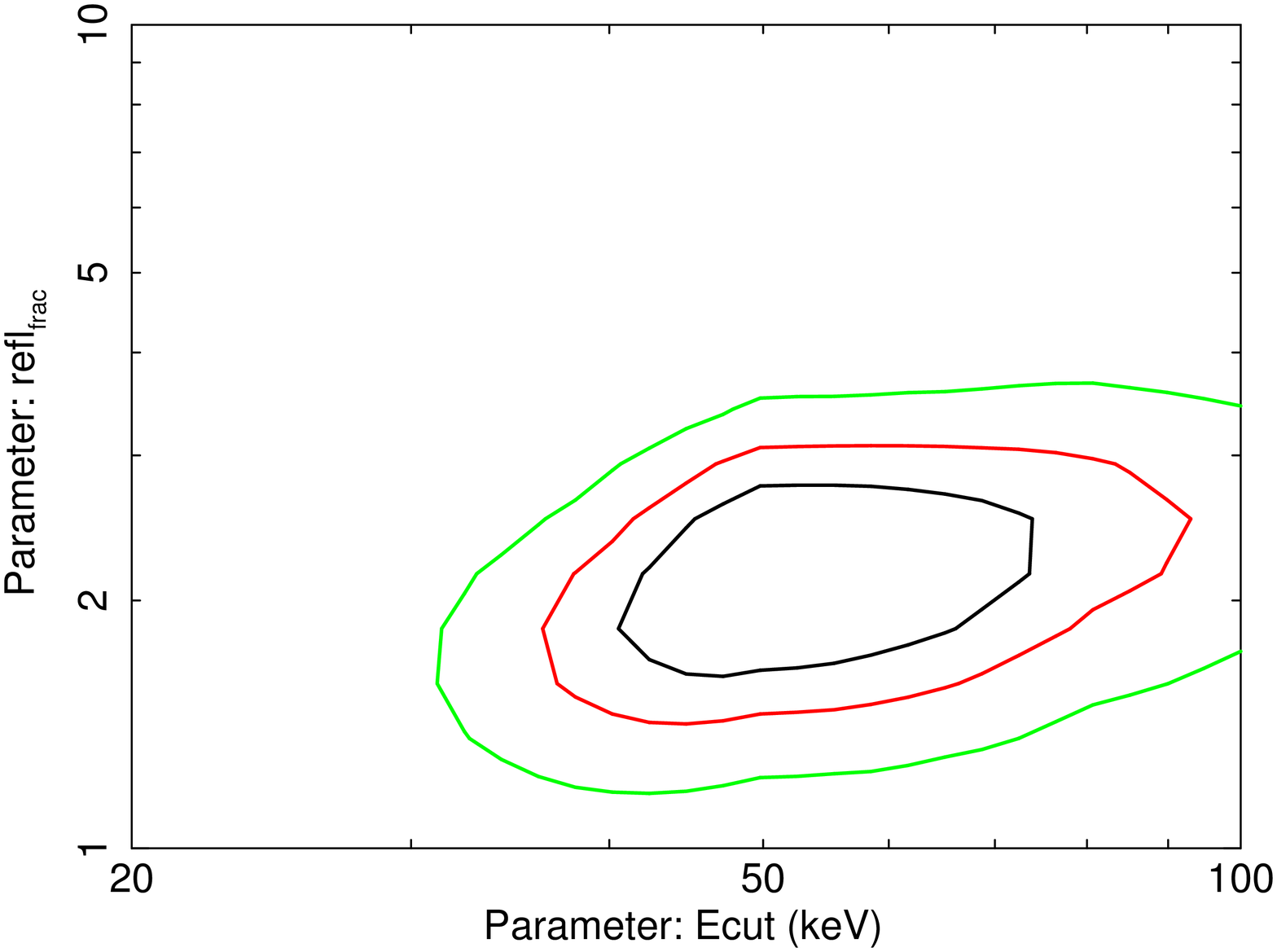}

\caption{\label{fig:contours} Contours of the goodness of fit, $\chi^{2}$,
around the best fitting values of two parameters: the power-law cut-off
energy and the reflection fraction. From inside out, the concentric
lines represent the $1\,\sigma$, $2\,\sigma$, and $3\,\sigma$ confidence
levels. Top: \emph{Suzaku} observation VII (best-fitting result at
domain boundary, $1\,\sigma$ contour not visible); middle: \emph{NuSTAR}
observation IX combined with overlap in \emph{Suzaku} VII; bottom:
\emph{NuSTAR} observation X with overlap in \emph{Suzaku} VII.}
\end{figure}
The spectral model of ionized reflection of a power-law continuum
suffers from degeneracies when fit to data of limited quality. Most
prominently, this is an issue for determining the parameters that
shape the Fe~K$\alpha$ line. For our fitting procedure we applied
a strategy (Section~\ref{sub:Reducing-degeneracies-in}) that revealed
strong correlations with flux for $\Gamma$ and $\log\xi$ (Section~\ref{sub:Correlations-with-Eddington}).
These parameters also shape the continuum of the spectra. Degeneracies
may also influence the other continuum parameters: the reflection
fraction and $E_{\mathrm{cutoff}}$. The correlations of these parameters
with flux are weaker, or are not quantified as precisely (Fig.~\ref{fig:ledd-xi}).
We investigate whether a strong degeneracy between the two parameters
influences our results in the three cases where we could measure $E_{\mathrm{cutoff}}$
(observations VII, IX, and XII; see Fig.~\ref{fig:ledd-xi}). Contours
of confidence intervals in $\chi^{2}$-space around the best fitting
values exhibit only minor asymmetries (Fig.~\ref{fig:contours}),
and no evidence of strong degeneracies is visible. For \emph{Suzaku}
observation VII, the best fit lies on the domain boundary of $E_{\mathrm{cutoff}}$.
Otherwise, our fit results are well localised. Furthermore, when we
determined the uncertainties in the fit parameters, we took any asymmetries
into account, such that our reported errors include the full extent
of the contours in Fig.~\ref{fig:contours}.

\bsp	% typesetting comment 
\label{lastpage}
\end{document}